\documentclass[12pt,a4paper,fleqn]{article}

\usepackage[DIV12]{typearea}
\usepackage{amsmath}
\usepackage{amssymb}
\usepackage{amsfonts}
\usepackage{amscd}
\usepackage{graphicx}
\usepackage[footnotesize]{caption2}
\usepackage{mcite}
\usepackage{mathrsfs}
\usepackage{bbm}

\newcommand{\D}{\ensuremath{\mathrm{d}}}

\DeclareMathOperator{\tr}{Tr}

\newcommand{\eVdist}{\kern-0.06667em}

\newcommand{\gev}{{\,\text{Ge}\eVdist\text{V\/}}}
\newcommand{\tev}{{\,\text{Te}\eVdist\text{V\/}}}

\newcommand{\ps}{\text{\sc ps}}
\newcommand{\GG}{\text{\sc gg}}
\newcommand{\vev}[1]{\ensuremath{\langle #1\rangle}}
\newcommand{\sm}{\text{\sc sm}}

\allowdisplaybreaks[1]

\begin{document}

\begin{titlepage}
\renewcommand{\thefootnote}{\alph{footnote}}

\begin{flushright}
DESY 05-238
\end{flushright}

\vspace*{1.0cm}

\renewcommand{\thefootnote}{\fnsymbol{footnote}}

\begin{center}
{\Large\bf Squarks and Sleptons between Branes and Bulk}

\vspace*{1cm}
\renewcommand{\thefootnote}{\alph{footnote}}

\textbf{
Wilfried Buchm\"uller\footnote[1]{Email: \texttt{wilfried.buchmueller@desy.de}},
J\"orn Kersten\footnote[2]{Email: \texttt{joern.kersten@desy.de}}
and
Kai Schmidt-Hoberg\footnote[3]{Email: \texttt{kai.schmidt.hoberg@desy.de}}
}
\\[5mm]

Deutsches Elektronen-Synchrotron DESY, 22603 Hamburg, Germany
\end{center}

\vspace*{1cm}

\begin{abstract}
\noindent
We study gaugino-mediated supersymmetry breaking in a six-dimensional $SO(10)$
orbifold GUT model where quarks and leptons are mixtures of brane and
bulk fields.  The couplings of bulk matter fields to the supersymmetry
breaking brane field have to be suppressed in order to avoid large
FCNCs.  We derive bounds on the soft supersymmetry breaking parameters
and calculate the superparticle mass spectrum.  If the gravitino is the
LSP, the $\tilde\tau_1$ or the $\tilde\nu_{\tau\mathrm{L}}$ turns out to
be the NLSP, with characteristic signatures at future colliders and in
cosmology.
\end{abstract}

\end{titlepage}
\newpage

\section{Introduction}
With the start of the LHC and the possible observation of superparticles
approaching, there is an increasing interest in specific predictions for
their mass spectrum, which results from the interplay
between fermion mass models and models for supersymmetry breaking.
Supersymmetric orbifold GUTs 
\cite{Kawamura:1999nj,*Kawamura:2000ev,*Altarelli:2001qj,*Hall:2001pg,%
*Hebecker:2001wq,*Asaka:2001eh,*Hall:2001xr}
are attractive candidates for
unified theories explaining the masses and mixings of fermions.
Features such as the doublet-triplet splitting and the absence of
dimension-five operators for proton decay, which are difficult to
realise in four-dimensional grand unified theories, are easily obtained.
Given the higher-dimensional setup with various branes, the mechanism of
supersymmetry breaking involves in general bulk as well as brane fields.

Following this rationale, we consider an $SO(10)$ theory in six
dimensions, proposed in \cite{Asaka:2003iy}, in combination with
gaugino-mediated SUSY breaking \cite{Kaplan:1999ac,Chacko:1999mi}.  The
orbifold compactification of the two extra dimensions has four fixed
points or ``branes''.  On three of them, three quark-lepton generations
are localised.  The Standard Model leptons and down-type quarks are
linear combinations of these localised fermions
and a partial fourth generation living in the bulk.  This leads to the
observed large neutrino mixings.  On the fourth brane, we assume a
gauge-singlet field $S$ to develop an $F$-term vacuum expectation value
(vev) breaking SUSY.  As the gauge and Higgs fields propagate in
the bulk, they feel the effects of SUSY breaking.  Thus, gauginos and
Higgs scalars obtain soft masses.  The soft masses and trilinear
couplings of
the scalar quarks and leptons approximately vanish at the
compactification scale.  Non-zero values are generated by the running to
low energies, which leads to a realistic superparticle mass spectrum.
If the gravitino is the lightest superparticle (LSP), it can be the
dominant component of the dark matter.  The next-to-lightest
superparticle (NLSP) is then a scalar tau or a scalar neutrino, which is
consistent with constraints from big bang nucleosynthesis.

In the next section, we will describe the orbifold model and the
couplings needed for SUSY breaking.  Subsequently, we will demonstrate
that the presence of extra matter fields in the bulk leads to severe
problems with flavour-changing neutral currents (FCNCs)
unless the couplings of these fields to the SUSY-breaking field $S$ are
suppressed.
Using na\"\i ve dimensional analysis (NDA), we derive upper bounds on
the unknown couplings of the theory and thus on the non-vanishing soft
masses, $\mu$ and $B\mu$ at high energy.  Finally, we calculate the
low-energy superparticle mass spectrum.  A realistic spectrum requires
the soft Higgs masses to satisfy bounds which are slightly stronger than
those estimated by NDA.

\section{The Orbifold GUT Model}
We consider an $N=1$
supersymmetric $SO(10)$ gauge theory in six dimensions 
compactified on the orbifold $T^2/\left(
      {\mathbbm Z}_2 \times {\mathbbm Z}_2^\prime  \times 
{\mathbbm Z}_2^{\prime\prime} \right)$ \cite{Asaka:2003iy}.
 The theory has four fixed points,
$O_\text{\sc i}$, $O_{\ps}$, $O_{\GG}$ and $O_\text{fl}$, located at
the corners of a ``pillow'' corresponding to the two compact
dimensions.  At $O_\text{\sc i}$ the full $SO(10)$ survives, 
whereas at the other fixed points, $O_{\ps}$,
$O_{\GG}$ and $O_\text{fl}$,  
$SO(10)$ is broken to its three GUT 
subgroups \mbox{${ G}_{\ps}={ SU(4)}\times { SU(2)} \times
  { SU(2)}$}, ${ G}_{\GG}={ SU(5)}\times {
  U(1)}_X$ and flipped $SU(5)$,
\mbox{${ G}_\text{fl}={ SU(5)'}\times { U(1)'}$},
respectively. The intersection of these GUT groups yields the
Standard Model group with an additional $U(1)$ factor, ${
    G}_{\sm '}= { SU(3)}\times { SU(2)} \times { U(1)}_Y
  \times { U(1)}_{X}$, as unbroken gauge symmetry below the
compactification scale, which we identify with the GUT scale.

The field content of the theory is strongly constrained by requiring
the cancellation of bulk and brane anomalies.
The brane fields are the three {\bf 16}-plets
$\psi_i$, $i=1,2,3$.  The bulk contains
six {\bf 10}-plets, $H_1,\dots, H_6$, and four {\bf 16}-plets, 
$\Phi, \Phi^c, \phi, \phi^c$, as hypermultiplets.  Vevs
of $\Phi$ and $\Phi^c$ break the surviving 
${U(1)}_{B-L}$.  The electroweak gauge group is broken by expectation
values of the doublets contained in $H_1$ and $H_2$.
  
We choose the parities of $\phi,\phi^c$ and $H_5,H_6$ such that their
zero modes are 
\begin{align}
\label{zeromodes}
  L = \left( \begin{array}{l} 
      \nu_4 \\ e_4
    \end{array} \right)\;, \quad  L^c = \left( \begin{array}{l} 
      n^c_4 \\ e^c_4 
    \end{array}\right)\;, \quad G^c_5 = d^c_4\;, \quad G_6 = d_4\;.
\end{align}
These zero modes act as a fourth generation of down (s)quarks
and (s)leptons and mix with the three generations of brane fields.  We
allocate the three sequential {\bf 16}-plets to the three branes where
$SO(10)$ is broken to its three GUT subgroups, placing $\psi_1$ at
$O_{\GG}$, $\psi_2$ at $O_\text{fl}$ and $\psi_3$ at $O_{\ps}$.  The three
``families'' are then separated by distances large compared to the
cutoff scale $\Lambda$. Hence, they can only have diagonal Yukawa
couplings with the bulk Higgs fields. The brane fields, however, can mix
with the bulk zero modes without suppression. As these mixings take place
only among left-handed leptons and right-handed down-quarks, we
obtain a characteristic pattern of mass matrices.

The model has the minimal amount of supersymmetry in six dimensions,
corresponding to $N=2$ extended supersymmetry in four dimensions.
The breaking to $N=1$
supersymmetry at the four-dimensional fixed points is achieved
by the ${\mathbbm Z}_2$-symmetry.
Soft SUSY-breaking terms are generated by gaugino mediation
\cite{Kaplan:1999ac,Chacko:1999mi}.  We place the gauge-singlet chiral
superfield $S$, which acquires a non-vanishing vev for its $F$-term
component, at the fixed point $O_\text{\sc i}$.  Supersymmetry is then
fully broken and the breaking 
can be communicated to gauge, Higgs and other bulk
fields by direct interactions.
The MSSM squarks and sleptons that live on branes can obtain soft
SUSY-breaking masses via loop contributions through the bulk, which are
negligible here, and via renormalisation group running.
To study the scalar masses and mixings we first have to discuss
all couplings which can lead to mass terms.

\subsection{The Superpotential}

The superpotential determines the SUSY-conserving mass terms 
and Yukawa couplings.
The allowed terms are restricted by $R$-invariance and an additional 
$U(1)_{\tilde{X}}$ symmetry with the charge assignments given in 
Tab. \ref{tab:charge}.
Starting from the six-dimensional theory, the effective 
four-dimensional fields are obtained by integrating out
the two extra dimensions.
This leads to a volume factor between the original six-dimensional fields
and the properly normalised fields we use here, $\Phi =\sqrt{V} \Phi_6$.

The most general brane superpotential without
the singlet field $S$ is given in \cite{Asaka:2003iy}.
All zero modes which have not been given in Eq.~\eqref{zeromodes}
can be found
in \cite{Asaka:2002nd}.
Since the fields $\psi_i$ and $\phi$ have the same quantum numbers,
they are combined to the quartet $\psi_\alpha = (\psi_i,\phi)$.
When the bulk fields are replaced by their zero modes, only $9$ 
of the $23$ terms appearing 
in the superpotential remain.
They are given by
\begin{align} \label{eq:WZeroModes}
W =& \; M^d H_5 H_6 + M^l_\alpha \psi_\alpha \phi^c 
      + \frac{1}{2}h^{(1)}_{\alpha \beta}\psi_\alpha
     \psi_\beta H_1 + \frac{1}{2}h^{(2)}_{\alpha \beta}\psi_\alpha \psi_\beta H_2
     + f_\alpha \Phi \psi_\alpha H_6  \nonumber \\
     &{}+ \frac{h^{N}_{\alpha \beta}}{2 \Lambda} \psi_\alpha \psi_\beta \Phi^c \Phi^c
      +\frac{g_\alpha^d}{\Lambda}\Phi^c \psi_\alpha H_5 H_1 
      + f^D \Phi^c \Phi^c H_3
      + f^G \Phi \Phi H_4 \;.
\end{align}
\begin{table}
  \centering
  \renewcommand{\arraystretch}{1.1}
  \begin{tabular}{|c||c|c|c|c|c|c|c|c|c|c|c|c|c|}
    \hline
     & $H_1$ & $H_2$ & $\Phi^c$ & $H_3$ & $\Phi$ & 
       $H_4$ & $\psi_i$ & $\phi^c$ & $\phi$ & $H_5$ & $H_6$ & $S$ \\
    \hline \hline
$R$ & 0 & 0 & 0 & 2 & 0 & 2 & 1 & 1 & 1 & 1 & 1 & 0 \\
    \hline 
$\tilde{X}$ & -2a & -2a & -a & 2a & a & -2a & a & -a & a & 2a & -2a & 0\\
    \hline
  \end{tabular}
  \caption{Charge assignments for the symmetries $U(1)_R$ 
           and $U(1)_{\tilde{X}}$
    \label{tab:charge}}
\end{table}

Consider now terms which involve the supersymmetry breaking singlet field $S$.
We want the brane Lagrangian to yield gaugino masses, since they
cannot be generated radiatively when starting from a vanishing mass at the
compactification scale.
Therefore, the source brane Lagrangian coupling the zero modes of the
gauge fields to the chiral field on the
source brane takes the form 
\begin{align}
\mathscr{L}_S    \supset \frac{g_4^2 h}{4 \Lambda}\,\int \text{d}^2\theta 
                 \,S \,W^\alpha W_\alpha 
                 + \text{h.c.}\;,
\end{align}
where $g_4$ is the four-dimensional gauge coupling and $h$ is a
dimensionless coupling.
From this equation and the ordinary kinetic term 
$\frac{1}{4}\int \text{d}^2\theta \, W^\alpha W_\alpha + \text{h.c.}$
for the gauge 
fields we conclude that
$S$ must have $U(1)_{\tilde{X}}$- and 
$R$-charge $0$ to leave the
Lagrangian invariant.
Therefore, terms respecting all the symmetries including
$U(1)_{\tilde{X}}$ are simply given by
\begin{align}
	\mathscr{L}_S \propto \int \text{d}^2\theta \, \frac{S}{\Lambda} \,W 
	+ \text{h.c.} \;,
\end{align}
where $W$ is the superpotential given above and where we only
keep those terms of $W$ which are at most cubic in the fields.
Note that in addition $\psi_\alpha$ has to be replaced by $\phi$,
since the matter fields $\psi_i$ cannot have direct couplings to the
source brane.
Moreover, we are interested only in terms which are non-zero
when replacing the fields by their zero modes.
A \textbf{16}-plet $\psi$ of $SO(10)$ is written in standard notation as
$\psi = (q,u^c,e^c,l,d^c,n^c).$
When setting the chiral field $S$ to its vev $F_S$, the scalar components
of the superfields remain, whereas the fermionic components are projected out.
When furthermore the \textbf{16}-plets $\Phi,\Phi^c$ 
acquire a vev $\vev{\Phi} = \vev{\Phi^c} = v_N \sim M_\mathrm{GUT}$
leading to the spontaneous breakdown of $U(1)_{B-L}$
we obtain (cf. Eq.~\eqref{zeromodes})
\begin{align} \label{eq:WSoft}
\mathscr{L}_S \supset - 
     \int \text{d}^2 \theta \; \frac{S}{\Lambda} 
      \left(\tilde{M}^d \tilde{d}_4^c \tilde{d}_4 + 
     \tilde{M}^l_4 \tilde{l}_4 \tilde{l}_4^c 
     \right) + \text{h.c.} \,.
\end{align}
Additional terms involving the heavy fields $\Phi,\Phi^c$ have been dropped.
Note that $U(1)_{B-L}$ is a subgroup of the local symmetry 
$U(1)_Y \times U(1)_X$.  The vevs $\vev{\Phi} = \vev{\Phi^c}$ break
$U(1)_X \times U(1)_{\tilde{X}}$ to a $U(1)$ subgroup.  As discussed in
\cite{Asaka:2002nd}, the superpotential \eqref{eq:WZeroModes} then
yields masses of order $M_\mathrm{GUT}$ for unwanted colour triplets
contained in $\Phi,\Phi^c,H_3$ and $H_4$.  In this way, the unification
of gauge couplings is maintained.

\subsection{The K\"ahler Potential}
In addition, soft mass terms can arise from the K\"ahler potential.
We assume the global $U(1)_{\tilde{X}}$ symmetry to be only approximate
and allow for explicit breaking here.
This is necessary in order to obtain a $\mu$-term, which is not allowed in the
superpotential, since the combination $H_1 H_2$ is not invariant under
$U(1)_{\tilde{X}}$.  Besides, an explicit breaking of
$U(1)_{\tilde X}$ is in fact required in order to avoid Goldstone
bosons.
Terms which result in non-negligible effects have to involve fields
which acquire a large vev in order to compensate for the suppression by the
cutoff scale $\Lambda$.
In our case, large vevs are acquired by $\Phi$, $\Phi^c$ and $S$.
We find that all terms without the singlet field $S$
do not contribute to any soft masses but merely give 
corrections to the kinetic terms.
Concentrating on the terms involving $S$, we do not consider terms 
with heavy fields that have no influence on low-energy physics.
In terms of the zero modes, the relevant part of the K\"ahler potential is
\begin{align}
\label{eq:softla}
	\mathscr{L}_S \supset&
	-\int\text{d}^4 \theta \,\bigg\{ \frac{S^\dagger}{\Lambda}
	\left(a  H_2 H_1^c  + b_1 H_1^{c\dagger} H_1^c 
                            + b_2 H_2^\dagger H_2\right) + \text{h.c.} 
\nonumber\\ 
& \hspace{20mm}
	+ \frac{1}{\Lambda^{2}} S^\dagger S \left[c_1 H_1^{c \dagger}H_1^c 
        + c_2 H_2^{\dagger}H_2
	+ \left(d H_2 H_1^c + \text{h.c.}\right) \right] \bigg\} \nonumber \\
&{} - \int \text{d}^4 \theta\,\bigg\{ \frac{e_i}{\Lambda}
              S^\dagger  B_i^\dagger B_i + \text{h.c.}
              + \frac{e_i^\prime}{\Lambda^2} S^{\dagger}S \,
              B_i^\dagger B_i \bigg\} \;,
\end{align}
yielding an effective $\mu$-term, soft Higgs masses, a $B\mu$-term and soft
masses for all other bulk fields.  $B_i$  ($i=1,\dots$) stands for any
bulk matter field except $H_{1,2}$.
Although the $\mu$-term itself is not a soft term, it is generated only
after the breaking of supersymmetry via the
Giudice-Masiero mechanism \cite{Giudice:1988yz}.  Note that there 
would be no electroweak symmetry breaking without the breaking of SUSY
and hence no massive (s)particles at the electroweak scale.

To see the contributions to the soft masses explicitly,
we express the Lagrangian 
by component fields, plugging in the 
$F$-term vev $F_S$ and the scalar vev $v_N$.
Furthermore, we employ the equations of motion for the auxiliary
fields and assume real couplings for simplicity.
Concentrating on the fourth generation and on the Higgs fields, 
this results in the following scalar mass terms:
\begin{align}
\label{lmsoft}
\mathscr{L}_S \supset& - \frac{F_S^{\dagger}F_S }{\Lambda^2}  
            \Big[ \left(a^2 + b_1^2 + c_1\right)\tilde{h}_1^{c\dagger} \tilde{h}_1^c 
             + \left(a^2 + b_2^2 + c_2\right)\tilde{h}_2^{\dagger} \tilde{h}_2 
             \nonumber \\
             &\hspace{18mm}+ \left(\vphantom{a_1^2}a\,(b_1 + b_2) +d\right)\tilde{h}_1^c\tilde{h}_2 + \text{h.c.} \Big] 
      \nonumber \\ 
&- \frac{F_S^{\dagger}F_S }{\Lambda^2}
              \Big[ \left(e_d^2 + e_d^\prime\right)\tilde{d}_4^\dagger \tilde{d}_4 
               + \left(e_{d^c}^2 + e_{d^c}^{\prime}\right)\tilde{d}_4^{c \dagger} 
                 \tilde{d}_4^c 
               +\left(e_l^2 + e_l^\prime\right) \tilde{l}_4^\dagger \tilde{l}_4 
               +\left(e_{l^c}^2 + e_{l^c}^\prime\right) \tilde{l}_4^{c \dagger} 
          \tilde{l}_4^c 
                            \Big] 
               \nonumber \\ 
& -\frac{F_S}{\Lambda} 
     \Big[ \tilde{M}^d \tilde{d}_4^c \tilde{d}_4 + 
     \tilde{M}^l_4 \tilde{l}_4 \tilde{l}_4^c 
     \Big] +\text{h.c.}
     \;,
\end{align}
where we have included the contribution from Eq.~\eqref{eq:WSoft} in the
last line.  We denote the components of a chiral multiplet by
$(\tilde{\phi},\phi,F_\Phi)$, with 
$\Phi = H_1^c, H_2, d_4, d_4^c, l_4, l_4^c$.
Note that the Higgs mass contribution proportional to $a^2$ is supersymmetric
and hence the soft Higgs masses are given by the terms proportional
to $(b_{1,2}^2 + c_{1,2})$.
		 
We assume that there are no sizable contributions to the scalar masses
from $D$-terms, which can arise when a gauged $U(1)$ symmetry is broken
or when there are soft SUSY breaking terms which lift a
$D$-flat direction in the scalar potential
\cite{Drees:1986vd,*Kolda:1995iw}.

\section{The Scalar Mass Matrices and FCNCs}
It is well known that in order to avoid flavour-changing neutral
currents (FCNCs), the squark and slepton mass matrices have to be
approximately diagonal in a basis where quark and lepton mass matrices
are diagonal.  In the following, we shall analyse this question for our
orbifold GUT model.  We have seen in the previous section that only the
scalars of the fourth generation, which are very heavy, obtain soft
masses, since they are bulk fields.  We will
demonstrate now that this leads to soft masses for the light scalars,
too.  At the compactification scale, we integrate out the heavy degrees
of freedom to obtain an effective theory with three generations.  This
requires diagonalising the mass matrices, and the corresponding
transformations transmit SUSY-breaking effects from the fourth to the
light generations.

From the zero mode superpotential \eqref{eq:WZeroModes}, one obtains the
mass terms
\begin{equation} \label{eq:WMass}
	W \supset
	d_\alpha m^d_{\alpha\beta} d^c_\beta +
	e^c_\alpha m^e_{\alpha\beta} e_\beta +
	n^c_\alpha m^D_{\alpha\beta} \nu_\beta +
	u^c_i m^u_{ij} u_j +
	\frac{1}{2} n^c_i M_{ij} n^c_j \;,
\end{equation}
where $m^d$, $m^e$ and $m^D$ are $4\times 4$ mass matrices of the form
\begin{equation} \label{eq:matrix-structure}
	m =
	\left(\begin{array}{cccc}
	 \mu_1 & 0 & 0 & \widetilde{\mu}_1 \\
	 0 & \mu_2 & 0 & \widetilde{\mu}_2 \\
	 0 & 0 & \mu_3 & \widetilde{\mu}_3 \\
	 \widetilde{M}_1 & \widetilde{M}_2 & \widetilde{M}_3 & \widetilde{M}_4
	\end{array} \right) .
\end{equation}
Here $\mu_i, \widetilde{\mu}_i \sim v$ and 
$\widetilde{M}_i \sim M_\mathrm{GUT}$.  While $\mu_i$ and
$\widetilde{\mu}_i$ have to be hierarchical, we assume no hierarchy
between the $\widetilde{M}_i$.
We have neglected corrections of order $\mathcal{O}(v_N/\Lambda)$.
For simplicity, we assume all matrices to be real.
The up-type quark and Majorana mass matrices $m^u$ and $M$ are diagonal
$3\times3$ matrices, since the corresponding fields do not have partners
in the bulk.

The mass matrices $m$ can be brought to the block-diagonal form
\begin{equation}
	m' = U_4^\dagger m V_4 =
	\begin{pmatrix} \widehat{m} & 0 \\ 0 & \widetilde{M} \end{pmatrix}
	+ \mathcal{O}\biggl(\frac{v^2}{\widetilde{M}}\biggr)
\end{equation}
by the transformation
\begin{align} \label{eq:BlockDiag}
	e \to e' = V_4^\dagger e \quad , \quad
	e^c \to e^{c\prime} = e^c \, U_4 \;,
\end{align}
where we have chosen the charged leptons for concreteness.
Here $U_4$ and $V_4$ are the unitary matrices
\begin{subequations}
\begin{align}
  U_4 &= \left(\begin{array}{cccc} 1&0&0&
      \frac{\mu_1\widetilde{M}_1 +
        \widetilde{\mu}_1\widetilde{M}_4}{\widetilde{M}^2} \\[0.5ex]
      0&1&0& \frac{\mu_2\widetilde{M}_2 +
        \widetilde{\mu}_2\widetilde{M}_4}{\widetilde M^2} \\[0.5ex]
      0&0&1& \frac{\mu_3\widetilde{M}_3 +
        \widetilde{\mu}_3\widetilde{M}_4}{\widetilde M^2} \\[0.5ex]
      -\frac{\mu_1 \widetilde{M}_1 + \widetilde{\mu}_1
        \widetilde{M}_4}{\widetilde{M}^2} & -\frac{\mu_2
        \widetilde{M}_2 + \widetilde{\mu}_2
        \widetilde{M}_4}{\widetilde{M}^2} & -\frac{\mu_3
        \widetilde{M}_3 + \widetilde{\mu}_3
        \widetilde{M}_4}{\widetilde{M}^2}& 1
    \end{array}\right)
  + \mathcal{O}\biggl(\frac{v^2}{\widetilde{M}^2}\biggr) \;,
  \\
  V_4 &= \left(\begin{array}{cccc}
      \frac{\widetilde{M}_4}{\widetilde{M}_{14}} & 0 &
      -\frac{\widetilde{M}_1\,
        \widetilde{M}_{23}}{\widetilde{M}\,\widetilde{M}_{14}} &
      \frac{\widetilde{M}_1}{\widetilde{M}}
      \\[0.9ex]
      0 &
      \frac{\widetilde{M}_3}{\widetilde{M}_{23}} &
      \frac{\widetilde{M}_2\, \widetilde
        M_{14}}{\widetilde{M}\,\widetilde{M}_{23}} &
      \frac{\widetilde{M}_2}{\widetilde{M}}
      \\[0.9ex]
      0 &
      -\frac{\widetilde{M}_2}{\widetilde{M}_{23}} &
      \frac{\widetilde{M}_3\,
        \widetilde{M}_{14}}{\widetilde{M}\,\widetilde{M}_{23}} &
      \frac{\widetilde M_3}{\widetilde{M}}
      \\[0.9ex]
      -\frac{\widetilde{M}_1}{\widetilde{M}_{14}}
      & 0 & -\frac{\widetilde{M}_4\,
        \widetilde{M}_{23}}{\widetilde{M}\,\widetilde{M}_{14}} &
      \frac{\widetilde M_4}{\widetilde{M}}
    \end{array}\right)
\end{align}
\end{subequations}
with $\widetilde{M} = \sqrt{\sum_\alpha \widetilde{M}_\alpha^2}$ and
$\widetilde{M}_{\alpha\beta} = \sqrt{\widetilde{M}_\alpha^2+\widetilde{M}_\beta^2}$
\cite{Buchmuller:2004eg}.
The transformation $V_4$ contributes to the desired large mixing between the
left-handed leptons.  $U_4$, on the other hand, is close to the unit
matrix, so that there is only small mixing among the right-handed
fields.  Note that the
situation is reversed in the down-quark sector, where the right-handed
fields are strongly mixed while the left-handed ones are not.

The SUSY-conserving charged slepton mass matrices are
$m_{e_\mathrm{L}}^2 = m^{e\dagger} m^e$ and
$m_{e_\mathrm{R}}^2 = m^e m^{e\dagger}$.  In addition, there are the
soft masses $m_{\tilde e_\mathrm{L}}^2$ etc.\ with non-zero 44-element.
Among them is the matrix $m_{\tilde e_\mathrm{LR}}^2$, which arises from
Eq.~\eqref{eq:WSoft} and mixes $\tilde l_4$ and $\tilde l_4^c$, but it
can be neglected for our purposes.  For the complete mass matrices we
use the notation
$m_{\tilde e_\mathrm{L},\mathrm{tot}}^2 = m_{e_\mathrm{L}}^2 +
 m_{\tilde e_\mathrm{L}}^2$ etc.
Under the transformation \eqref{eq:BlockDiag}, they change to
\begin{subequations}
\begin{align}
	m_{\tilde e_\mathrm{L},\mathrm{tot}}^{\prime2} &=
	V_4^\dagger \, m_{e_\mathrm{L}}^2 \, V_4 +
	V_4^\dagger \, m_{\tilde e_\mathrm{L}}^2 \, V_4
\nonumber\\
&=
	\begin{pmatrix}
	 \widehat m^\dagger \widehat m & 0 \\ 0 & \widetilde M^2
	\end{pmatrix} +
	V_4^\dagger
	 \begin{pmatrix} 0 & 0 \\ 0 & m^2_{\tilde l_{4\mathrm{L}}} \end{pmatrix} V_4 +
	\mathcal{O}(v^2)
	 \begin{pmatrix}
	  \frac{v^2}{\widetilde M^2} & 1 \\ 1 & 1
	 \end{pmatrix} ,
\\
	m_{\tilde e_\mathrm{R},\mathrm{tot}}^{\prime2} &=
	U_4^\dagger \, m_{e_\mathrm{R}}^2 \, U_4 +
	U_4^\dagger \, m_{\tilde e_\mathrm{R}}^2 \, U_4
\nonumber\\
&=
	\begin{pmatrix}
	 \widehat m \widehat m^\dagger & 0 \\
	 0 & \widetilde M^2 + m^2_{\tilde l_{4\mathrm{R}}}
	\end{pmatrix} +
	\mathcal{O}\biggl( \frac{v^3}{\widetilde{M}},
	 \frac{v \,m^2_{\tilde l_{4\mathrm{R}}}}{\widetilde{M}} \biggr) \;,
\end{align}
\end{subequations}
where the fourth-generation soft masses are denoted by $m^2_{\tilde
l_{4\mathrm{L}}}$ and $m^2_{\tilde l_{4\mathrm{R}}}$, in analogy to
those of the first three generations, although both $l_4$ and $l_4^c$
are $SU(2)_\mathrm{L}$ doublets.
The matrices are block-diagonal up to rotations by angles of order
$v^2/\widetilde M^2$ or smaller, which can safely be neglected.
The soft masses of the light ``right-handed'' sleptons are highly
suppressed.  However, this is not true for their ``left-handed''
counterparts, whose $3\times3$ mass matrix is given by
\begin{equation}
	(m_{\tilde e_\mathrm{L},\mathrm{tot}}^{\prime2})_{ij} =
	(\widehat m^\dagger \widehat m)_{ij} +
	(V_4)_{4i} (V_4)_{4j} \, m_{\tilde l_{4\mathrm{L}}}^2 =
	(\widehat m^\dagger \widehat m)_{ij} +
	(\widehat m^2_{\tilde e_\mathrm{L}})_{ij} \;.
\end{equation}

The light fermion mass matrix $\widehat m$ is diagonalised by a second
change of basis, 
\begin{equation}
	m_\mathrm{diag} = V_\mathrm{CKM} \, \widehat m \, \widehat V \;.
\end{equation}
In the approximation $\mu_1=\mu_2=0$, the transformation matrix
$\widehat V$ is known explicitly \cite{Buchmuller:2004eg},
\begin{align} \label{eq:vhat}
  \widehat V = \left( \begin{array}{ccc} -\frac{\widetilde{M}_2\,
        \widetilde{M}_4}{\widetilde{M}_{12}\, \widetilde{M}_{14}} 
& \frac{\widetilde{M}_1 \left( \widetilde{\mu}_3
          \widetilde{M}_3\, \widetilde{M}_4-\mu_3 \widetilde M_{124}^2
        \right)}{\bar\mu_3\, \widetilde{M}\, \widetilde{M}_{12}\, 
     \widetilde{M}_{14}} &
      -\frac{\widetilde{\mu}_3}{\bar\mu_3}\frac{\widetilde{M}_1}{\widetilde{M}_{14}}
      \\[1ex]
     \frac{\widetilde{M}_1\, \widetilde{M}_3}{\widetilde{M}_{12}\, \widetilde{M}_{23}}
      & \frac{\widetilde{M}_2 \left( \widetilde{\mu}_3 \widetilde{M}_{123}^2
      -\mu_3\widetilde{M}_3\, \widetilde{M}_4 \right)}{\bar\mu_3\, \widetilde{M}\,
        \widetilde{M}_{12}\, \widetilde{M}_{23}} &
      -\frac{\mu_3}{\bar\mu_3}\frac{\widetilde{M}_2}{\widetilde{M}_{23}}
      \\[1.1ex]
      \frac{\widetilde{M}_1\, \widetilde{M}_2\, \widetilde{M}}{\widetilde{M}_{12}\,
        \widetilde{M}_{14}\, \widetilde{M}_{23}} 
       & -\frac{\widetilde{\mu}_3 \widetilde{M}_1^2\,
        \widetilde{M}_3+\mu_3\widetilde{M}_2^2\, \widetilde{M}_4}{\bar\mu_3\, 
        \widetilde{M}_{12}\, \widetilde{M}_{14}\,
        \widetilde{M}_{23}} & -\frac{\widetilde{\mu}_3}{\bar\mu_3}\frac{\widetilde{M}_4\,
        \widetilde{M}_{23}}{\widetilde{M}\, \widetilde{M}_{14}} + \frac{\mu_3}{\bar\mu_3}
      \frac{\widetilde{M}_3\, \widetilde{M}_{14}}{\widetilde{M}\, \widetilde{M}_{23}}
    \end{array} \right)
\end{align}
up to a rotation of the second and third generation by a
small angle given by the ratio of the 23- and 33-elements of
$\widehat V^\dagger \, \widehat m^\dagger \widehat m \, \widehat V$,
$\Theta_R \simeq
 (\widetilde{\mu}_1^2+\widetilde{\mu}_2^2)/\bar\mu_3^2 \ll 1$.
In Eq.~\eqref{eq:vhat}, we have defined
$\widetilde{M}_{\alpha\beta\gamma}^2 =
 \widetilde M_\alpha^2 + \widetilde M_\beta^2 + \widetilde M_\gamma^2$
and
\begin{equation} \label{eq:b3}
  \bar\mu^2_3 = \widetilde{\mu}_3^2 \left( 1- \frac{\widetilde{M}_4^2}{\widetilde{M}^2}\right)
  + \mu_3^2 \left(1- \frac{\widetilde{M}_3^2}{\widetilde{M}^2}\right) - 2\mu_3\widetilde{\mu}_3
  \frac{\widetilde{M}_3\widetilde{M}_4}{\widetilde{M}^2} \;.
\end{equation}
We finally obtain for the charged slepton mass matrix in the basis where
the charged fermion mass matrix is diagonal
{
\setlength{\mathindent}{0pt}
\begin{align*}
&
	\widehat V^\dagger \,
	(\widehat m^\dagger \widehat m + \widehat m^2_{\tilde e_\mathrm{L}}) \,
	\widehat V =
\\
&
	\begin{pmatrix}
	 0 & 0 & 0 \\
	 0 &
	 \frac{\mu_3^2 \widetilde M_{12}^2}{\bar\mu_3^2 \widetilde M^2}
	  \left( \widetilde\mu_1^2 + \widetilde\mu_2^2 + m^2_{\tilde l_{4\mathrm{L}}} \right) &
	 \frac{\mu_3 \widetilde M_{12}
	  \left( \widetilde\mu_3 \widetilde M_{123}^2 -
	   \mu_3 \widetilde M_3 \widetilde M_4 \right)}
	  {\bar\mu_3^2 \widetilde M^3}
	  \left( \widetilde\mu_1^2 + \widetilde\mu_2^2 + m^2_{\tilde l_{4\mathrm{L}}} \right) \\
	 0 &
	 \frac{\mu_3 \widetilde M_{12}
	  \left( \widetilde\mu_3 \widetilde M_{123}^2 -
	   \mu_3 \widetilde M_3 \widetilde M_4 \right)}
	  {\bar\mu_3^2 \widetilde M^3}
	  \left( \widetilde\mu_1^2 + \widetilde\mu_2^2 + m^2_{\tilde l_{4\mathrm{L}}} \right) \! &
	 \bar\mu_3^2 + 
     \frac{\left(\widetilde{\mu}_3 \widetilde{M}_{123}^2 -\mu_3\widetilde{M}_{3}
     \widetilde{M}_{4}\right)^2}{\bar{\mu}_3^2 \widetilde{M}^4}
	 m^2_{\tilde l_{4\mathrm{L}}} +
	 \mathcal{O}\bigl(
	  \frac{\widetilde\mu_1^2+\widetilde\mu_2^2}{\bar\mu_3^2} \bigr)
	\end{pmatrix} .
\end{align*}
}%
The non-zero off-diagonal elements are of similar size as the diagonal
elements, unless $m_{\tilde l_{4\mathrm{L}}} \ll \bar\mu_3 \sim m_\tau$.
Numerically, we find that the same is true for the 12- and 13-entries,
if $\mu_1$ and $\mu_2$ are non-zero.  This leads to unacceptably large
FCNCs in the lepton sector.  The situation in the down quark sector is
analogous.  We expect this problem to be generic in higher-dimensional
theories with mixing between bulk and brane matter fields as long as the
bulk fields can couple to the hidden sector (cf.\ e.g.\ 
\cite{Buchmuller:2005jr}).  In the following, we shall
assume that soft masses for bulk matter fields, contrary to the bulk
Higgs fields, are strongly suppressed, i.e.\ 
$m_{\tilde l_{4\mathrm{L,R}}} \simeq m_{\tilde d_{4\mathrm{L,R}}} \simeq 0$.  Within the present
framework of orbifold GUTs, the coupling of brane and bulk fields cannot
be understood dynamically.

\section{Constraints from Na\"\i ve Dimensional Analysis}

The couplings of the brane field $S(x)$ to bulk fields $B(x,y)$ can
be constrained by na\"\i ve dimensional analysis \cite{Chacko:1999hg}.
For this purpose, one rewrites the relevant part of the six-dimensional Lagrangian
\begin{equation} \label{eq:LDCanonical}
	\mathscr{L}_{B,S} =
	\mathscr{L}_\mathrm{bulk}(B(x,y)) +
	\delta^{2}(y-y_S) \, \mathscr{L}_S(B(x,y),S(x))
\end{equation}
in terms of dimensionless fields $\hat B(x,y)$ and $\hat S(x)$,
and the cutoff $\Lambda$, up to which the theory should be valid,
\begin{equation} \label{eq:LDDimless}
	\mathscr{L}_{B,S} =
	\frac{\Lambda^6}{\ell_6/C} \,
	 \mathscr{\hat L}_\mathrm{bulk}(\hat B(x,y)) +
	 \delta^{2}(y-y_S) \, \frac{\Lambda^4}{\ell_4/C} \,
	 \mathscr{\hat L}_S(\hat B(x,y),\hat S(x)) \;,
\end{equation}
where $\ell_6 = 128 \pi^3$ and $\ell_4 = 16\pi^2$.
Here $y_S$ corresponds to the extra-dimensional coordinates of the
brane where the singlet field $S(x)$ resides, $y_S=(0,0)$.
The factor $C$ accounts for the multiplicity of fields in loop diagrams
for a non-Abelian gauge group.
The rescaling of chiral bulk and brane superfields reads  
\begin{equation}
B(x,y) = \left( \frac{\Lambda^{4}V}{\ell_6/C} \right)^{1/2}\hat B(x,y)
\quad , \quad
S(x) = \left( \frac{\Lambda^2}{\ell_4/C} \right)^{1/2} \hat S(x) \;.
\label{eq:phiAndphiHat}
\end{equation}
Note the additional factor of $\sqrt{V}$ for the bulk field due to the
proper normalisation.

The combination $C/\ell_D$ gives the typical geometrical suppression of
loop diagrams.  This suppression is cancelled by the factors $\ell_6/C$
and $\ell_4/C$ in front of the Lagrangians $\mathscr{\hat L}$ in
Eq.~\eqref{eq:LDDimless}.  Consequently, all loops will be of the same order
of magnitude, provided that all couplings are $\mathcal{O}(1)$.  Thus,
according to the NDA recipe the effective six-dimensional theory remains
weakly coupled up to the cutoff $\Lambda$, if the dimensionless
couplings in Eq.~\eqref{eq:LDDimless} are smaller than one.

Let us apply NDA to the part of the brane Lagrangian 
giving rise to Higgs and Higgsino masses, the first two lines of
Eq.~\eqref{eq:softla}.
Using Eq.~\eqref{eq:phiAndphiHat}, we obtain
\begin{multline}
\mathscr{L}_S \supset -\frac{\Lambda^4}{\ell_4/C} \,\int 
              \frac{\text{d}^4 \theta}{\Lambda^2} \, \bigg\{ 
              \frac{V \Lambda^2 \sqrt{\ell_4 C}}{\ell_6} 
                \left(a \hat S^\dagger  \hat H_2 \hat H_1^c 
                 + b_1 \hat S^\dagger \hat H_1^{c\dagger} \hat H_1^c 
                            + b_2 \hat S^\dagger \hat H_2^\dagger \hat H_2  
                + \text{h.c.}\right) 
\\ 
                + \frac{V \Lambda^2 C}{\ell_6}\hat S^\dagger \hat S 
               \left[ c_1 \hat H_1^{c \dagger}\hat H_1^c 
                  + c_2 \hat H_2^{\dagger}\hat H_2 
                  + \bigl(d \hat H_2 \hat H_1^c 
                  + \text{h.c.}\bigr)\right] \bigg\} \;.
\end{multline}
The NDA requirement that all couplings be smaller than one implies the
following constraints on $a,b_{1,2},c_{1,2},d$:
\begin{subequations}
\begin{align}
              \frac{V \Lambda^2 \sqrt{\ell_4 C}}{\ell_6} \,(a,b_1,b_2) &< 1 \;,
\\
               (c_1,c_2,d) \, \frac{V \Lambda^2 C}{\ell_6} &< 1 \;.
\end{align}
\end{subequations}
Using $\ell_4 = 16\pi^2$ and $\ell_6 = 128\pi^3$, one then obtains upper
bounds on the couplings at the compactification scale,
\begin{subequations}
\begin{align}
	(a,b_1,b_2) &< \frac{32\pi^2}{V \Lambda^2 \sqrt{C}} \;,
\\
	(c_1,c_2,d) &< \frac{128\pi^3}{V \Lambda^2 C} \;.
\end{align}
\end{subequations}
These inequalities translate into upper bounds on the $\mu$- and $B\mu$-terms
and on the soft Higgs masses,
\begin{subequations}
\begin{align}
	\mu &= \frac{a F_S^\dagger}{\Lambda}
	\,<\, \frac{32\pi^2 F_S^\dagger} {V \sqrt{C}\Lambda^3} \;,
\\
	B\mu 
	&= \frac{\bigl(a\,(b_1 + b_2) + d\bigr) \, F_S^\dagger F_S}{\Lambda^2}
	\,<\,  \Bigl( 1 + \frac{16\pi}{V \Lambda^2} \Bigr) \, 
          \frac{128\pi^3 F_S^\dagger F_S}{V C \Lambda^4} \;,
\\
	(m^2_{\tilde{h}_2},m^2_{\tilde{h}_1}) 
	&= \frac{(c_2 +b_2^2,c_1 + b_1^2) \, F_S^\dagger F_S}{\Lambda^2}
	\,<\, \Bigl( 1 + \frac{8\pi}{V \Lambda^2} \Bigr) \, 
              \frac{128\pi^3 F_S^\dagger F_S}{V C \Lambda^4} \;.
\end{align}
\end{subequations}

Applying the NDA recipe to those terms of Eqs.~\eqref{eq:WSoft} and
\eqref{eq:softla} giving 
rise to soft superparticle masses we obtain
\begin{multline}
\mathscr{L}_S \supset -\frac{\Lambda^4}{\ell_4/C}\, \bigg[
	\int \frac{\text{d}^2\theta}{\Lambda} \,
	\frac{\sqrt{\ell_4/C}}{\ell_6/C} \Lambda^2 V \,
	\biggl( \frac{\tilde{M}^d}{\Lambda} \hat S \hat H_5 \hat H_6 + 
	\frac{\tilde{M}^l_4}{\Lambda} \hat S \hat\phi \hat\phi^c \biggr) 
	+ \text{h.c.} 
\\
	\qquad + \int \frac{\text{d}^4 \theta}{\Lambda^2} \,
	\frac{\sqrt{\ell_4/C}}{\ell_6/C} \Lambda^2 V \,
	\bigg\{ e_i  
	\hat S^\dagger  \hat B_i^\dagger \hat B_i + \text{h.c.}
	+ \frac{e_i^\prime}{\sqrt{\ell_4/C}} \hat S^{\dagger} \hat S \,
	\hat B_i^\dagger \hat B_i  \bigg\}\bigg] \;. 
\end{multline}
The resulting upper bounds on the masses can be found in Tab.\
\ref{tab:nda}.
We have also included the bound on the gaugino mass derived in
\cite{Buchmuller:2005rt} and the gravitino mass.

To be more explicit, we make assumptions about the values of the 
parameters involved.
The compactification scale is assumed to be of order the unification scale,
$V^{-1/2} = M_\mathrm{GUT} = 2.5 \cdot 10^{16}\gev$.
The cutoff $\Lambda$ is given by the six-dimensional Planck scale,
$\Lambda = M_6 = M_4^{1/2} V^{-1/4} = 2.4 \cdot 10^{17}\gev$.
We choose $C=C_2(G)$ for the group theory factor, which gives $C=8$ for
the gauge group $G=SO(10)$.
%
\begin{table}
  \centering
  \renewcommand{\arraystretch}{1.7}
  \begin{tabular}{|c||c|c|c|}
    \hline
 $m_{1/2}$    &     $ = \frac{g_4^2 h F_S}{2 \Lambda}$ 
              &  $< \frac{16 \pi^2 F_S}{\sqrt{C} V \Lambda^3}$&$< 1 \tev$ \\
    \hline  
$m_{\tilde d_{4\mathrm{RL}}}^2$  &  $= \frac{F_S}{\Lambda} \tilde{M}^d$
              &  $< \frac{32\pi^2 F_S}{\sqrt{C} V \Lambda^2}$ 
              &  $< (2 \cdot 10^{7}\tev)^2$\\
    \hline 
$m_{\tilde l_{4\mathrm{LR}}}^2$  & $= \frac{F_S}{\Lambda} \tilde{M}^l_4$
              &  $< \frac{32\pi^2 F_S}{\sqrt{C} V \Lambda^2}$ 
              &  $< (2\cdot 10^{7}\tev)^2$     \\
    \hline
$m_{\tilde d_{4\mathrm{L}}}^2$
 & $= (e_d^2 + e_d^\prime) \, \frac{F_S^\dagger F_S}{\Lambda^2}$
              &  $< (1+\frac{8\pi}{V\Lambda^2}) \,
                    \frac{128\pi^3 F_S^\dagger F_S}{CV \Lambda^4}$ 
              & $< (4  \tev)^2$ \\
    \hline
$m_{\tilde l_{4\mathrm{L}}}^2$
 & $= (e_l^2 + e_l^\prime) \, \frac{F_S^\dagger F_S}{\Lambda^2}$
              &  $< (1+\frac{8\pi}{V\Lambda^2}) \,
                    \frac{128\pi^3 F_S^\dagger F_S}{CV \Lambda^4}$ 
              & $< (4  \tev)^2$ \\
    \hline
$\mu$         &  $= a \frac{F_S^\dagger}{\Lambda}$ 
              &  $< \frac{32\pi^2 F_S^\dagger} {\sqrt{C} V \Lambda^3}$ 
              &  $< 2   \tev$  \\ 
    \hline
$(m^2_{\tilde{h}_2},m^2_{\tilde{h}_1})$ 
              & $= (c_2+b_2^2,c_1+b_1^2) \, \frac{F_S^\dagger F_S}{\Lambda^2}$
              &  $< (1+\frac{8\pi}{V\Lambda^2}) \,
                 \frac{128\pi^3 F_S^\dagger F_S}{C V \Lambda^4}$ 
              &  $< (4 \tev)^2$ \\    
    \hline
$B\mu$        & $= \bigl(a\,(b_1 + b_2) + d\bigr) \, \frac{F_S^\dagger F_S}{\Lambda^2}$
              &  $<(1+\frac{16\pi}{V\Lambda^2}) \,
                  \frac{128\pi^3 F_S^\dagger F_S}{C V \Lambda^4}$ 
              &  $< (5 \tev)^2$ \\
    \hline
$m_{3/2}$ & $=\frac{F_S}{\sqrt{3}M_4}$ & & = $100 \gev$ \\
    \hline 
  \end{tabular}
\caption{NDA constraints on mass parameters.  The numerical values are
 valid for $F_S = 4\cdot 10^{20} \gev^2$. The masses for the fields 
$l_4^c,d_4^c$ are analogous to those of $l_{4},d_{4}$.
}
\label{tab:nda}
\end{table}
%
This leads to the numerical values for the NDA bounds shown in the last
column of Tab.~\ref{tab:nda}.

\section{The Low-Energy Sparticle Spectrum}
Imposing $m_{\tilde l_{4\mathrm{L,R}}}=m_{\tilde d_{4\mathrm{L,R}}}=0$,
the boundary conditions
at the compactification scale are those of the usual gaugino mediation
scenario with bulk Higgs fields \cite{Chacko:1999mi},
\begin{subequations}
\begin{align}
	g_1 &= g_2 = g_3 = g \simeq \frac{1}{\sqrt{2}} \;,
\\
	M_1 &= M_2 = M_3 = m_{1/2} \;,
\\
	m_{\tilde \phi_\mathrm{L}}^2 =
	m_{\tilde \phi_\mathrm{R}}^2 &= 0
	\quad \text{for all squarks and sleptons } \tilde\phi \;,
\\
	A_{\tilde\phi} &= 0
	\quad \text{for all squarks and sleptons } \tilde\phi\;,
\\
	\mu, B\mu, m^2_{\tilde{h}_i} &\neq 0 \quad (i=1,2) \;,
\end{align}
\end{subequations}
where GUT charge normalisation is used for $g_1$.  The scalar mass
matrices then remain almost diagonal, so that FCNCs are suppressed.
We have neglected
small corrections to the scalar masses from gaugino loops
\cite{Kaplan:1999ac} as well as corrections to the gauge couplings from
brane-localised terms breaking the unified gauge symmetry.  For
$m^2_{\tilde{h}_i}=0$, these boundary conditions have previously been
considered in different contexts in \cite{Ellis:1984bm,*Inoue:1991rk}.
Upper limits on the non-vanishing
parameters are summarised in Tab.\ \ref{tab:nda}.  If we choose a
certain value for the universal gaugino mass $m_{1/2}$, this implies a
lower bound on the vev $F_S$ according to the first row of the table.
The choice is constrained by the lower
bound on the Higgs mass from LEP, $m_{h^0} > 114.4\gev$, because lighter
gauginos imply a lighter Higgs.  In addition, we require the gravitino
to be lighter than $100\gev$, which leads to an upper bound on $F_S$.

As a benchmark point for our discussion, we choose $m_{1/2} = 500\gev$,
$\tan\beta=10$ and $\text{sign}(\mu)=+1$, which yields
\begin{subequations}
\begin{align}
	2 \cdot 10^{20}\gev^2 < F_S &< 4 \cdot 10^{20}\gev^2 \;,
\\
	50\gev < m_{3/2} &< 100\gev \;.
\end{align}
\end{subequations}
We use the current best-fit value $m_t=172.7\gev$ \cite{CDF:2005cc} for
the top mass.  The values of $\mu$ and $B\mu$ are then determined by the
conditions for electroweak symmetry breaking.
We find that their numerical values at the compactification scale
are well below their NDA bounds.

In order to find the spectrum at low energy, we have to take into
account the running of the parameters.  We employ SOFTSUSY
\cite{Allanach:2001kg} for this purpose.  To obtain an analytical
understanding of the results, let us consider the one-loop renormalisation group
equations (RGEs) for the soft masses at the compactification scale
\cite{Inoue:1982pi,*Inoue:1983pp},
\begin{subequations}
\begin{align}
	16\pi^2 \frac{\D M_i^2}{\D t} &= 4 b_i g^2 m_{1/2}^2 \;,
\label{eq:MiDot}
\\
	16\pi^2 \frac{\D m_{\tilde q_{3\mathrm{L}}}^2}{\D t} &=
	-\frac{84}{5} g^2 m_{1/2}^2 + \frac{1}{5} g^2 \tr(Y m^2) + X_t + X_b
	\;,
\label{eq:m2q3LDot}
\\
	16\pi^2 \frac{\D m_{\tilde t_\mathrm{R}}^2}{\D t} &=
	-\frac{64}{5} g^2 m_{1/2}^2 - \frac{4}{5} g^2 \tr(Y m^2) + 2 X_t \;,
\label{eq:m2tRDot}
\\
	16\pi^2 \frac{\D m_{\tilde b_\mathrm{R}}^2}{\D t} &=
	-\frac{56}{5} g^2 m_{1/2}^2 + \frac{2}{5} g^2 \tr(Y m^2) + 2 X_b \;,
\\
	16\pi^2 \frac{\D m_{\tilde \tau_\mathrm{L}}^2}{\D t} &=
	-\frac{36}{5} g^2 m_{1/2}^2 - \frac{3}{5} g^2 \tr(Y m^2) + X_\tau\;,
\label{eq:m2StauLDot}
\\
	16\pi^2 \frac{\D m_{\tilde \tau_\mathrm{R}}^2}{\D t} &=
	-\frac{24}{5} g^2 m_{1/2}^2 + \frac{6}{5} g^2 \tr(Y m^2) +2X_\tau\;,
\label{eq:m2StauRDot}
\\
	16\pi^2 \frac{\D m^2_{\tilde{h}_1}}{\D t} &=
	-\frac{36}{5} g^2 m_{1/2}^2 - \frac{3}{5} g^2 \tr(Y m^2) + 3 X_b
	+ X_\tau \;,
\\
	16\pi^2 \frac{\D m^2_{\tilde{h}_2}}{\D t} &=
	-\frac{36}{5} g^2 m_{1/2}^2 + \frac{3}{5} g^2 \tr(Y m^2) + 3 X_t \;,
\label{eq:mh2Dot}
\end{align}
\end{subequations}
where $t = \ln\frac{\mu}{\mu_0}$ with the renormalisation scale $\mu$,
$b_i = (\frac{33}{5},1,-3)$ are the coefficients in the RGEs of the
gauge couplings,
\begin{equation}
	16\pi^2 \frac{\D g_i^2}{\D t} = 2 b_i g^4 \;,
\end{equation}
and
\begin{subequations}
\begin{align}
	X_t &= 2 y_t^2 m^2_{\tilde h_2}
	\simeq \frac{1}{2} \left( 1+\cot^2\beta \right) m^2_{\tilde h_2}\;,
\\
	X_b &=
	 2 y_b^2 \, m^2_{\tilde h_1}
	\simeq 5 \cdot 10^{-5} \left( 1+\tan^2\beta \right)
 	 m^2_{\tilde h_1} \;,
\\
	X_\tau &=
	 2 y_\tau^2 \, m^2_{\tilde h_1}
	\simeq 10^{-4} \left( 1+\tan^2\beta \right)
 	 m^2_{\tilde h_1} \;.
\end{align}
\end{subequations}
The numerical values in the previous equations represent the typical
orders of magnitude of the top, bottom and tau Yukawa couplings at high
energy.  We assume a not too large $\tan\beta$, so that $X_b$ and
$X_\tau$ are negligible at the GUT scale.  However, we will
see that $X_\tau$ can become relevant at lower energies.  The term
$\tr(Y m^2)$, often abbreviated by $S$, vanishes for universal scalar
masses but plays an important role in our case, if one of the soft Higgs
masses is sufficiently large.  At $M_\mathrm{GUT}$, it is given by
\begin{equation}
	\tr(Y m^2) = m^2_{\tilde h_2} - m^2_{\tilde h_1} \;.
\end{equation}
The RGEs
for the first and second generation scalar masses are obtained from the
above equations by omitting $X_t$, $X_b$ and $X_\tau$.  We do not list
the RGEs for $\mu$, $B\mu$ and the $A$-terms, since they are not
relevant for our discussion.  We will also use
\begin{subequations}
\begin{align}
	16\pi^2 \frac{\D (g_i^2 M_i^2)}{\D t} &=
	6 b_i g^4 m_{1/2}^2 \;,
\label{eq:RGEg2M2}
\\
	16\pi^2 \frac{\D \tr(Y m^2)}{\D t} &=
	\frac{66}{5} g^2 \tr(Y m^2) \;.
\label{eq:RGETrYm2}
\end{align}
\end{subequations}

\subsection{Gaugino Masses}
The 1-loop RGEs \eqref{eq:MiDot} for the gaugino masses do not depend on
the scalar masses, so that their low-energy values remain virtually
the same in all cases as long as we do not change $m_{1/2}$.
Numerically, we find
\begin{subequations}
\begin{align}
	M_1(M_Z) &\simeq 200\gev \;,
\\
	M_2(M_Z) &\simeq 380\gev \;,
\\
	M_3(M_Z) &\simeq 1200\gev \;.
\end{align}
\end{subequations}
To good approximation, the lightest neutralino is the bino and the
second-lightest one is the wino, unless $m^2_{\tilde{h}_2}$ is sizable.  In
the latter case, the electroweak symmetry breaking conditions lead to a
rather small $\mu$, so that there is
significant mixing between the neutralinos.

\subsection{Allowed Parameter Space for the Soft Higgs Masses}
\begin{figure}
\centering
\includegraphics{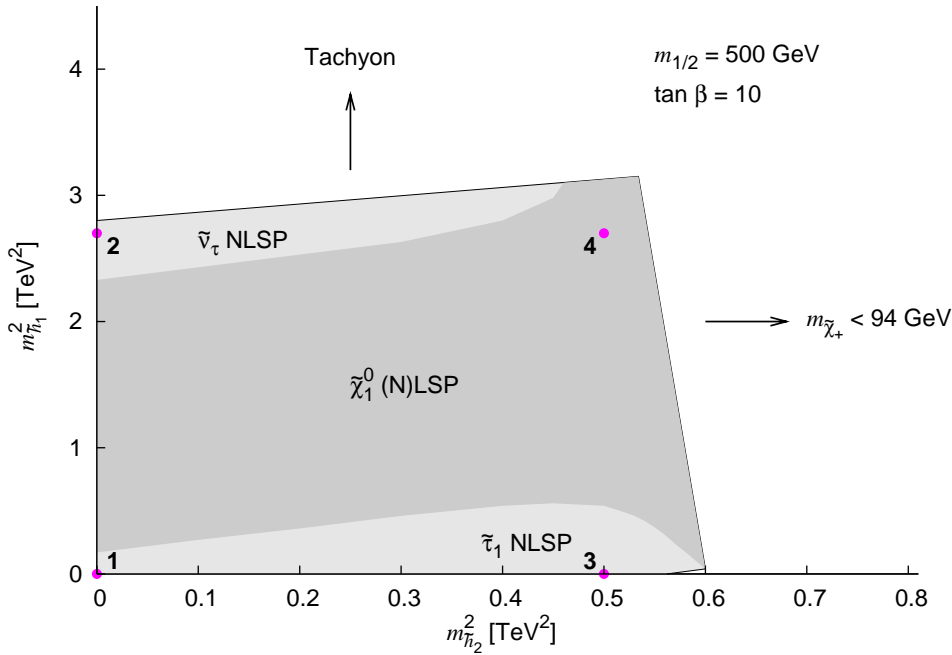}
\caption{Allowed region for the soft Higgs masses.  In the dark-gray
 area, a neutralino is lighter than all sleptons.  For the points marked
 by the coloured dots, the resulting superparticle mass spectrum is
 shown in Fig.~\ref{fig:LineSpectra}.} 
\label{fig:parameterspace}
\end{figure}

In addition to the constraints from NDA, there are phenomenological
limits on the soft masses $m_{\tilde{h}_i}^2$ at the compactification
scale, which turn out to be more restrictive.  The resulting allowed
region in parameter space is the gray-shaded area in
Fig.~\ref{fig:parameterspace}.

One constraint is that the running of the parameters
down to the weak scale must not produce tachyons.
For scalar masses which vanish at the compactification scale this means that
their $\beta$-function must not be positive there.%
\footnote{Strictly speaking, a scalar mass squared may arrive at a
 positive value at low energies even if its $\beta$-function is
 positive at the compactification scale.  We do not take
 this possibility into account, so that the constraints are
 conservative.}
The one-loop RGE \eqref{eq:m2StauLDot} for the left-handed sleptons gives the 
most restrictive constraint on $m_{\tilde{h}_1}^2$,
\begin{align} \label{eq:UpperBoundmh1}
m_{\tilde{h}_1}^2 < 12 \, m_{1/2}^2 + m_{\tilde{h}_2}^2 \;.
\end{align}

The upper bounds on $m_{\tilde{h}_2}^2$ are due to the experimental
limits on the superparticle masses \cite{Eidelman:2004wy}.  If the
initial value of $m_{\tilde{h}_2}^2$ is too large, this mass squared 
crosses zero at a rather low energy, so that its absolute value at the
electroweak scale is small.  Consequently, the $\mu$ parameter is also
small, leading to a Higgsino-like chargino with a mass below the current
limit of $94\gev$.  If we increased $m_{\tilde{h}_2}^2$ further, there
would be no successful electroweak symmetry breaking.  This limit on
$m_{\tilde{h}_2}^2$ is the relevant one for almost all values of  
$m_{\tilde{h}_1}^2$.  Only for very small $m_{\tilde{h}_1}^2$, the 
experimental requirement that the lighter stau be heavier than $86\gev$
becomes more restrictive.

For simplicity we only consider positive soft Higgs masses at the
compactification scale.  With negative soft masses,
it is possible to end up in the ``light Higgs window'' at the electroweak
scale, though only in a very narrow parameter range.
In this window with both $m_{h_0}$ and $m_A$ around 
$90$ -- $100\gev$ we can decrease $m_{1/2}$ significantly down to 
at least $250 \gev$.

\subsection{Dependence of the Spectrum on the Higgs Masses}
Due to the large effects of the strong interaction, the squark masses
experience the fastest running and end up around a TeV.  
The lighter stop mass runs more slowly due to $X_t$, which is always
sizable at lower energies, and reaches a value of about $800\gev$.
If all scalar soft masses vanish at the GUT scale, the
left-handed slepton masses change significantly in the beginning, but
afterwards the evolution flattens as $g_1^2 M_1^2$ and $g_2^2 M_2^2$
decrease (cf.\ Eq.~\eqref{eq:RGEg2M2}).  Hence, they reach intermediate
values between $300$ and $400\gev$ at low energies.  The flattening of the evolution is
even more pronounced for the right-handed slepton masses, since here it
depends only on $g_1^2 M_1^2$, which decreases faster than 
$g_2^2 M_2^2$.  As a consequence, these scalars remain lighter than the lightest
neutralino \cite{Kaplan:1999ac}, which is approximately the bino:
$m_{\tilde e_\mathrm{R}}(M_Z) \simeq 180\gev$,
$m_{\chi^0_1} \simeq 200\gev$.  For both slepton ``chiralities'', the
third generation is slightly lighter than the first two due to $X_\tau$.

\begin{figure}
\centering
\includegraphics{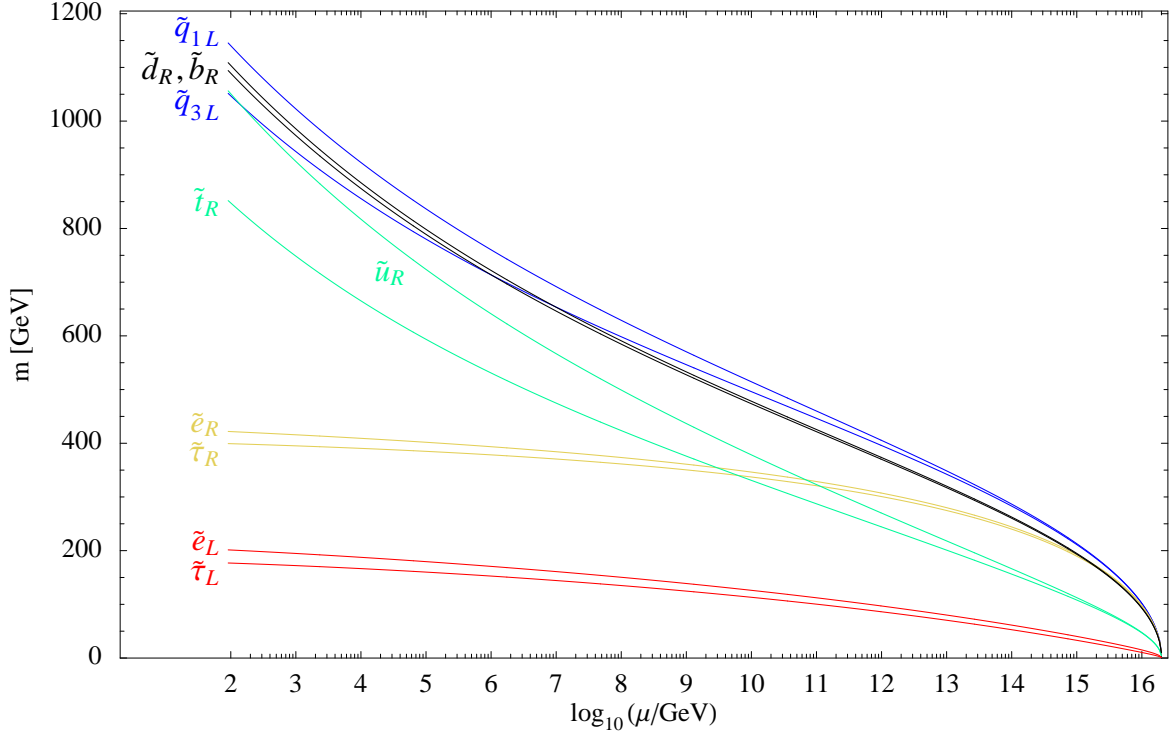}
\caption{Evolution of the scalar soft masses for
 $m^2_{\tilde{h}_1}=2.7\tev^2$, 
 $m^2_{\tilde{h}_2}=0$ (point 2 in Fig.~\ref{fig:parameterspace}),
 $m_{1/2}=500\gev$, $\tan\beta=10$ and $\text{sign}(\mu)=+1$ at
 $M_\mathrm{GUT}$.
}
\label{fig:RunningScalarMasses}
\end{figure}

For $m^2_{\tilde{h}_1} \neq m^2_{\tilde{h}_2}$, the term involving
$\tr(Y m^2)$ is non-vanishing and can lead to important changes
\cite{Lleyda:1993xf,Chacko:1999mi,Kaplan:2000av,Schmaltz:2000ei}.  We
shall first consider the case where it is negative 
($m^2_{\tilde{h}_1} > m^2_{\tilde{h}_2}$) and saturates the bound from
Eq.~\eqref{eq:UpperBoundmh1} (numerically, we find a slightly stronger
bound of $m^2_{\tilde{h}_1} - m^2_{\tilde{h}_2} < 2.7\tev^2$, which we
use here).  Then $|\tr(Y m^2)/m_{1/2}^2| \sim 10$, so that the first and
second terms on the r.h.s.\ of the RGEs \eqref{eq:m2q3LDot} --
\eqref{eq:m2StauRDot} can be of the same order of magnitude.
An example for the running of the scalar masses is shown in
Fig.~\ref{fig:RunningScalarMasses}.

The most drastic change occurs in the slepton spectrum.  For the largest
possible value of $\tr(Y m^2)$, the r.h.s.\ of Eq.~\eqref{eq:m2StauLDot}
vanishes exactly at the GUT scale.  It turns negative only at lower
energies due to the fast decrease of $|\tr(Y m^2)|$ (cf.\
Eq.~\eqref{eq:RGETrYm2}).  As a result, the left-handed sleptons remain
relatively light, with a low-energy mass below $200\gev$.  Contrary to
that, both terms in the RGE for the right-handed slepton masses are of
the same sign, leading to an unusually fast running near the GUT scale.
At lower energies, the evolution slows down quickly due to the fast
decrease of both $g_1^2 M_1^2$ and $|\tr(Y m^2)|$.  The resulting masses
are close to $400\gev$.  Thus, the NLSP is a sneutrino in this case
\cite{Kaplan:2000av}, with a slightly heavier stau $\tilde\tau_1$ due to
the $SU(2)_\mathrm{L}$ and $U(1)_\mathrm{Y}$ $D$-terms.

In the squark sector, large masses are generated again due to the strong
interaction.  At high energies, there is a significant cancellation in
the RGE \eqref{eq:m2tRDot} for the right-handed up-type squark masses,
while the contributions to the other squark mass RGEs add up.
Consequently, $m_{\tilde u_\mathrm{R}}$ and $m_{\tilde t_\mathrm{R}}$
run quite slowly until $|\tr(Y m^2)|$ has decreased sufficiently.
Afterwards, $m_{\tilde u_\mathrm{R}}$ runs faster and comes close to the
masses of the left-handed and right-handed down-type squarks at the
electroweak scale.

If $m^2_{\tilde{h}_1}$ is neither close to zero nor to its upper bound,
the running of the right-handed slepton masses is sufficiently enhanced
to lift them above the lightest neutralino mass.  At the same time, the
running of the left-handed slepton masses is damped weakly enough, so
that they are heavier than the lightest neutralino, too
\cite{Chacko:1999mi,Kaplan:2000av}.  A neutralino NLSP together with a
gravitino LSP heavier than a GeV is excluded by cosmology 
\cite{Fujii:2003nr,*Feng:2004mt,Ellis:2003dn,*Kawasaki:2004qu,*Cerdeno:2005eu}.
Therefore, this case is only viable if the neutralino is the LSP and the
gravitino is heavier.  This is possible, because we only have a lower
bound on the gravitino mass.  The corresponding region in parameter
space is marked by the dark-gray area in Fig.~\ref{fig:parameterspace}.
It grows for large values of $m^2_{\tilde{h}_2}$, since then mixing
additionally decreases the lightest neutralino mass.
A neutralino LSP is also often obtained if the compactification scale is
larger than the unification scale.  In this case, the running above
$M_\mathrm{GUT}$ tends to make the sleptons heavier than the lightest
neutralino \cite{Schmaltz:2000ei,Baer:2001ze}.

For $m^2_{\tilde{h}_2} > m^2_{\tilde{h}_1}$, $\tr(Y m^2)$ is positive.
Now the evolution of the right-handed slepton masses is slowed down by
the $\tr(Y m^2)$-term, while that of the left-handed masses is enhanced.
Consequently, the NLSP is the predominantly right-handed $\tilde\tau_1$,
with a mass of about
$100\gev$ for $m^2_{\tilde{h}_2} = 0.5\tev^2$ and $m^2_{\tilde{h}_1}=0$.
For these values, the masses of the left-handed sleptons are roughly
$350\gev$.

Since $m^2_{\tilde{h}_2}$ cannot be much larger than $0.5\tev^2$, the RGEs
for the squark masses are always dominated by the term proportional to
$g_3^2 M_3^2$.  Consequently, the low-energy masses are almost unchanged
compared to the case of vanishing soft scalar masses at the
compactification scale, except for $m_{\tilde q_{3\mathrm{L}}}$ and
$m_{\tilde t_\mathrm{R}}$, which decrease by up to $60\gev$ due to the
larger $X_t$.

\begin{figure}
\centering
\includegraphics[width=\textwidth]{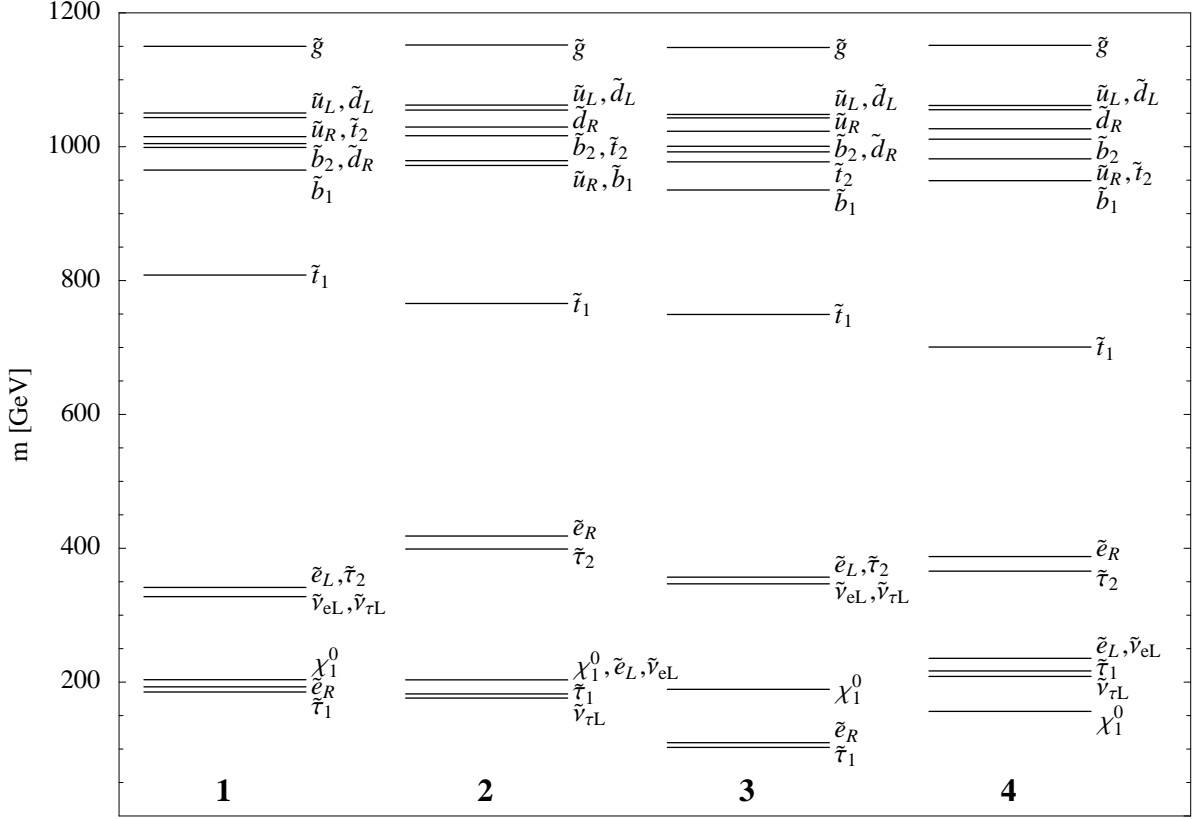}
\caption{
 Spectra of superparticle pole masses.  The numbers at the bottom
 correspond to the points in parameter space marked by the coloured dots
 in Fig.~\ref{fig:parameterspace}.  The high-energy boundary conditions
 for the soft Higgs masses were
 $m^2_{\tilde{h}_1}=m^2_{\tilde{h}_2}=0$ (point 1),
 $m^2_{\tilde{h}_1}=2.7\tev^2$, $m^2_{\tilde{h}_2}=0$ (point 2),
 $m^2_{\tilde{h}_1}=0$, $m^2_{\tilde{h}_2}=0.5\tev^2$ (point 3), and
 $m^2_{\tilde{h}_1}=2.7\tev^2$, $m^2_{\tilde{h}_2}=0.5\tev^2$ (point 4),
 respectively.  In all cases, we used $m_{1/2}=500\gev$, $\tan\beta=10$
 and $\text{sign}(\mu)=+1$.  As the first and second generation scalars
 are degenerate, only the first generation is listed in the figure.
 Particles with a mass difference of less than about $3\gev$ are
 represented by a single line.  The heavier neutralinos and the
 charginos have been omitted for better readability.
}
\label{fig:LineSpectra}
\end{figure}

As the dominant parts of the RGEs depend only on the difference
$m^2_{\tilde{h}_1} - m^2_{\tilde{h}_2}$, the same is true for the
spectrum to a good approximation.  The sum is only relevant for those
third-generation masses whose evolution is sensitive to the $X_i$,
most notably $m_{\tilde t_\mathrm{R}}$.
In Fig.~\ref{fig:LineSpectra}, we show the superparticle spectra that we
obtain at the four points in parameter space marked by the coloured dots
in Fig.~\ref{fig:parameterspace}.

\subsection{Dependence on the Gaugino Masses}
To a first approximation, varying the high-energy gaugino mass simply
leads to a rescaling of the scalar spectrum.  If $m_{1/2}$ is increased
while keeping the other soft masses fixed, the relative sizes of 
$\tr(Y m^2)$ and the $X_i$ decrease.  Hence, they become less important and
the spectrum comes closer to the one obtained in the minimal case of
vanishing scalar masses.

As mentioned before, the LEP bound on the lightest Higgs mass leads to a
lower bound on $m_{1/2}$.  Actually, with our benchmark value
$m_{1/2}=500\gev$ we obtain a Higgs mass slightly below $114\gev$ for
small soft masses.  However, the mass can easily be pushed beyond the
bound by raising the top mass by about $1.5\gev$ above its present
best-fit value of $172.7\gev$.
Furthermore, a non-zero mass $m^2_{\tilde{h}_1}$ also causes
an increase of the Higgs mass.  If $m^2_{\tilde{h}_1}$ takes the maximal
value allowed by Eq.~\eqref{eq:UpperBoundmh1},
a unified gaugino mass of slightly less than $400\gev$ is compatible
with the LEP bound (for $m_t=172.7\gev$).

\subsection[Dependence on $\tan\beta$]{Dependence on $\boldsymbol{\tan\beta}$}
The influence of $\tan\beta$ on the results is also rather straightforward to
understand. As to the RGEs, it only enters in the parameters $X_t$,
$X_b$ and $X_\tau$, which play a role in the evolution of the
third-generation soft masses.  Hence, a change of $\tan\beta$ leads to a
change of the mass splitting between this generation and the first two.

If $\tan\beta$ is significantly smaller than 10, the value used in our
benchmark scenario, $X_t$ increases. Consequently, $\tilde t_\mathrm{R}$
and $\tilde q_{3\mathrm{L}}$ become slightly lighter. On the other hand,
$X_b$ and $X_\tau$ are negligible now, so that the inter-generation mass
splitting in the slepton and right-handed down-type squark sector
becomes tiny.  The Higgs mass bound leads to severer restrictions now.  If
$\tan\beta<8$, raising the top mass to the maximal value of $175.6\gev$
allowed by experiment no longer yields $m_{h^0}>114.4\gev$ for 
$m^2_{\tilde{h}_1}=0$.  If $\tan\beta<6$, the bound is violated even for
maximal $m_t$ and $m^2_{\tilde{h}_1}$, i.e.\ a gaugino mass larger than
$500\gev$ is required.

\begin{figure}
\centering
\includegraphics{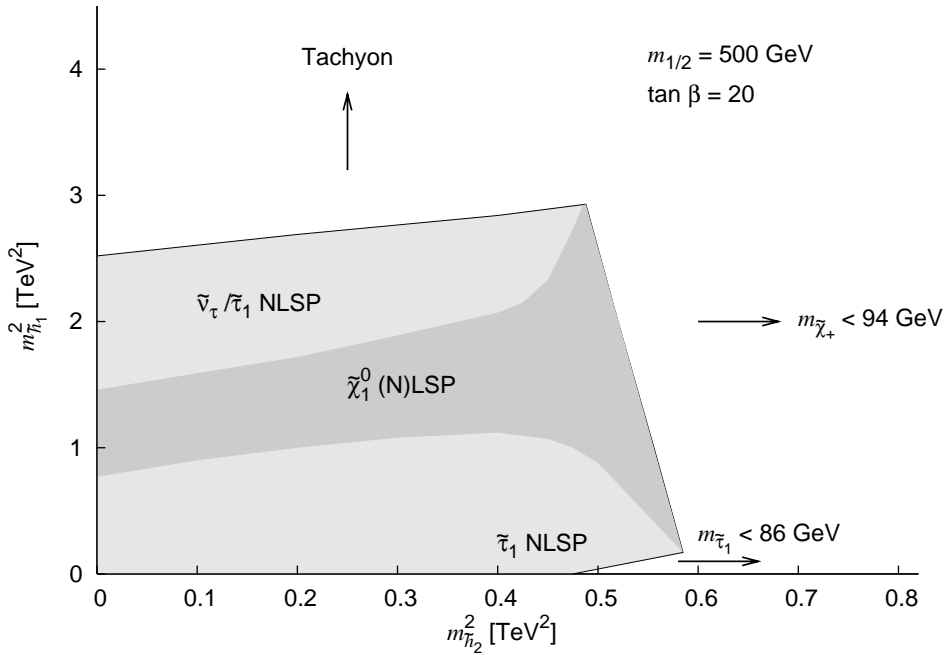}
\caption{Allowed region for the soft Higgs masses for $\tan\beta=20$.
 In the dark-gray area, a neutralino is lighter than all sleptons.} 
\label{fig:parameterspace2}
\end{figure}

For larger values of $\tan\beta$, $X_b$ and $X_\tau$ become more
important.
Nevertheless, the impact of the former parameter on the RG evolution
remains subdominant compared to that of the strong interaction.  Hence,
its increase only causes a larger splitting between 
$m_{\tilde d_\mathrm{R}}$ and $m_{\tilde b_\mathrm{R}}$, but does not
lead to any new restrictions.  In contrast, the lighter stau mass
decreases a lot faster at lower energies due to the larger $X_\tau$.  On
the one hand, this increases the parameter space region where the
$\tilde\tau_1$ is lighter than the neutralinos, as shown in
Fig.~\ref{fig:parameterspace2} for $\tan\beta=20$. 
On the other hand, the soft scalar
masses have to satisfy severer upper bounds in order to avoid tachyons
and a too light stau.
As $\tan\beta$ increases beyond 20, mixing causes an additional decrease
of $m_{\tilde\tau_1}$, as the off-diagonal term in the mass matrix,
$\left( m^2_{\tilde\tau} \right)_{12} \simeq -v \mu y_\tau$, becomes
comparable to the diagonal entries.  
For $\tan\beta=25$, the region of parameter space where the neutralino
is lighter than the $\tilde\tau_1$ almost vanishes.
For $\tan\beta=35$, the model is only viable if all soft scalar masses
vanish at the GUT scale, and for $\tan\beta>35$ the lighter stau mass
always lies below its experimental limit.%
\footnote{Valid points in parameter space may exist for negative soft
 Higgs masses, but even in this case the allowed region is rather small.
}

These problems are alleviated for heavier gauginos.  In order to
obtain a viable model with $\tan\beta=50$, one requires
$m_{1/2} \gtrsim 850\gev$, if all other soft masses vanish.  If they are
non-zero, the gaugino mass has to be even larger.  In the resulting
spectrum, only one stau is relatively light, while the remaining
superparticle masses lie above $300\gev$.  As the lower bound on the
gravitino mass rises with $m_{1/2}$, the gravitino may become heavier
than the stau, which is excluded by cosmology. 

We conclude that the model favours $10 \lesssim \tan\beta \lesssim 25$.
For values far outside this range, the phenomenological bounds on the
soft masses are much more restrictive than the NDA limits, which appears
unnatural.

\section{Conclusions}
We have discussed gaugino-mediated SUSY breaking in a six-dimensional
$SO(10)$ orbifold GUT model where quarks and leptons are mixtures of brane
and bulk fields.  The couplings of bulk matter fields to the SUSY
breaking gauge singlet brane field have to be suppressed in order to
avoid large FCNCs.  The compatibility of the SUSY breaking mechanism and
orbifold GUTs with brane and bulk matter fields is a generic problem
which requires further studies.  We have also determined bounds on the
SUSY breaking parameters by na\"\i ve dimensional analysis, which turn
out not to restrict the phenomenologically allowed parameter regions.

The parameters relevant for the superparticle mass spectrum are the
universal gaugino mass, the
soft Higgs masses, $\tan\beta$ and the sign of $\mu$.  We have analysed
their impact on the spectrum and determined the region in parameter
space that results in a viable phenomenology.  The model favours
moderate values of $\tan\beta$ between about 10 and 25.  The gaugino
mass at the GUT scale should not be far below $500\gev$ in order to
satisfy the LEP bound on the Higgs mass.  Typically, the lightest
neutralino is bino-like with a mass of $200\gev$, and the gluino mass is
about $1.2\tev$.  Either the right-handed or the left-handed sleptons
can be lighter than the neutralinos.  The corresponding region in
parameter space grows with $\tan\beta$.  In this region, the gravitino
is the LSP with a mass around $50\gev$.  The $\tilde\tau_1$ or the
$\tilde\nu_{\tau\mathrm{L}}$ is the NLSP.
A sneutrino NLSP has the advantage that constraints from big bang
nucleosynthesis and the cosmic microwave background are less stringent
\cite{Fujii:2003nr,*Feng:2004mt}.  For a stau NLSP, on the other hand,
there exists
the exciting possibility that its decays may lead to the discovery of
the gravitino in future collider experiments
\cite{Buchmuller:2004rq,*Brandenburg:2005he}.

\section*{Acknowledgements}
We would like to thank Ben Allanach, Dirk Br\"ommel,
Koichi Hamaguchi, Tilman Plehn, Michael Ratz and Peter Zerwas
for valuable discussions.
This work has been supported by the ``Impuls- und Vernetzungsfonds'' of
the Helmholtz Association, contract number VH-NG-006.

\frenchspacing

\bibliography{GauginoMediation}

\providecommand{\bysame}{\leavevmode\hbox to3em{\hrulefill}\thinspace}
\begin{mcbibliography}{10}

\bibitem{Kawamura:1999nj}
Y.~Kawamura, Prog. Theor. Phys. \textbf{103} (2000), 613 [hep-ph/9902423]\relax
\relax
\bibitem{Kawamura:2000ev}
Y.~Kawamura, Prog. Theor. Phys. \textbf{105} (2001), 999 [hep-ph/0012125]\relax
\relax
\bibitem{Altarelli:2001qj}
G.~Altarelli, F.~Feruglio, Phys. Lett. \textbf{B511} (2001), 257
  [hep-ph/0102301]\relax
\relax
\bibitem{Hall:2001pg}
L.~J. Hall, Y.~Nomura, Phys. Rev. \textbf{D64} (2001), 055003
  [hep-ph/0103125]\relax
\relax
\bibitem{Hebecker:2001wq}
A.~Hebecker, J.~March-Russell, Nucl. Phys. \textbf{B613} (2001), 3
  [hep-ph/0106166]\relax
\relax
\bibitem{Asaka:2001eh}
T.~Asaka, W.~Buchm{\"u}ller, L.~Covi, Phys. Lett. \textbf{B523} (2001), 199
  [hep-ph/0108021]\relax
\relax
\bibitem{Hall:2001xr}
L.~J. Hall, Y.~Nomura, T.~Okui, D.~R. Smith, Phys. Rev. \textbf{D65} (2002),
  035008 [hep-ph/0108071]\relax
\relax
\bibitem{Asaka:2003iy}
T.~Asaka, W.~Buchm{\"u}ller, L.~Covi, Phys. Lett. \textbf{B563} (2003), 209
  [hep-ph/0304142]\relax
\relax
\bibitem{Kaplan:1999ac}
D.~E. Kaplan, G.~D. Kribs, M.~Schmaltz, Phys. Rev. \textbf{D62} (2000), 035010
  [hep-ph/9911293]\relax
\relax
\bibitem{Chacko:1999mi}
Z.~Chacko, M.~A. Luty, A.~E. Nelson, E.~Ponton, JHEP \textbf{01} (2000), 003
  [hep-ph/9911323]\relax
\relax
\bibitem{Asaka:2002nd}
T.~Asaka, W.~Buchm{\"u}ller, L.~Covi, Phys. Lett. \textbf{B540} (2002), 295
  [hep-ph/0204358]\relax
\relax
\bibitem{Giudice:1988yz}
G.~F. Giudice, A.~Masiero, Phys. Lett. \textbf{B206} (1988), 480\relax
\relax
\bibitem{Drees:1986vd}
M.~Drees, Phys. Lett. \textbf{B181} (1986), 279\relax
\relax
\bibitem{Kolda:1995iw}
C.~F. Kolda, S.~P. Martin, Phys. Rev. \textbf{D53} (1996), 3871
  [hep-ph/9503445]\relax
\relax
\bibitem{Buchmuller:2004eg}
W.~Buchm{\"u}ller, L.~Covi, D.~Emmanuel-Costa, S.~Wiesenfeldt, JHEP \textbf{09}
  (2004), 004 [hep-ph/0407070]\relax
\relax
\bibitem{Buchmuller:2005jr}
W.~Buchm{\"u}ller, K.~Hamaguchi, O.~Lebedev, M.~Ratz, hep-ph/0511035\relax
\relax
\bibitem{Chacko:1999hg}
Z.~Chacko, M.~A. Luty, E.~Ponton, JHEP \textbf{07} (2000), 036
  [hep-ph/9909248]\relax
\relax
\bibitem{Buchmuller:2005rt}
W.~Buchm{\"u}ller, K.~Hamaguchi, J.~Kersten, Phys. Lett. \textbf{B632} (2006),
  366 [hep-ph/0506105]\relax
\relax
\bibitem{Ellis:1984bm}
J.~R. Ellis, C.~Kounnas, D.~V. Nanopoulos, Nucl. Phys. \textbf{B247} (1984),
  373\relax
\relax
\bibitem{Inoue:1991rk}
K.~Inoue, M.~Kawasaki, M.~Yamaguchi, T.~Yanagida, Phys. Rev. \textbf{D45}
  (1992), 328\relax
\relax
\bibitem{CDF:2005cc}
{CDF Collaboration}, {D0 Collaboration}, {Tevatron Electroweak Working Group},
  hep-ex/0507091\relax
\relax
\bibitem{Allanach:2001kg}
B.~C. Allanach, Comput. Phys. Commun. \textbf{143} (2002), 305
  [hep-ph/0104145]\relax
\relax
\bibitem{Inoue:1982pi}
K.~Inoue, A.~Kakuto, H.~Komatsu, S.~Takeshita, Prog. Theor. Phys. \textbf{68}
  (1982), 927\relax
\relax
\bibitem{Inoue:1983pp}
K.~Inoue, A.~Kakuto, H.~Komatsu, S.~Takeshita, Prog. Theor. Phys. \textbf{71}
  (1984), 413\relax
\relax
\bibitem{Eidelman:2004wy}
Particle Data Group, S.~Eidelman et~al., Phys. Lett. \textbf{B592} (2004),
  1\relax
\relax
\bibitem{Lleyda:1993xf}
A.~Lleyda, C.~Mu{\~n}oz, Phys. Lett. \textbf{B317} (1993), 82
  [hep-ph/9308208]\relax
\relax
\bibitem{Kaplan:2000av}
D.~E. Kaplan, T.~M.~P. Tait, JHEP \textbf{06} (2000), 020
  [hep-ph/0004200]\relax
\relax
\bibitem{Schmaltz:2000ei}
M.~Schmaltz, W.~Skiba, Phys. Rev. \textbf{D62} (2000), 095004
  [hep-ph/0004210]\relax
\relax
\bibitem{Fujii:2003nr}
M.~Fujii, M.~Ibe, T.~Yanagida, Phys. Lett. \textbf{B579} (2004), 6
  [hep-ph/0310142]\relax
\relax
\bibitem{Feng:2004mt}
J.~L. Feng, S.~Su, F.~Takayama, Phys. Rev. \textbf{D70} (2004), 075019
  [hep-ph/0404231]\relax
\relax
\bibitem{Ellis:2003dn}
J.~R. Ellis, K.~A. Olive, Y.~Santoso, V.~C. Spanos, Phys. Lett. \textbf{B588}
  (2004), 7 [hep-ph/0312262]\relax
\relax
\bibitem{Kawasaki:2004qu}
M.~Kawasaki, K.~Kohri, T.~Moroi, Phys. Rev. \textbf{D71} (2005), 083502
  [astro-ph/0408426]\relax
\relax
\bibitem{Cerdeno:2005eu}
D.~G. Cerde{\~n}o, K.-Y. Choi, K.~Jedamzik, L.~Roszkowski, R.~Ruiz~de Austri,
  hep-ph/0509275\relax
\relax
\bibitem{Baer:2001ze}
H.~Baer, A.~Belyaev, T.~Krupovnickas, X.~Tata, Phys. Rev. \textbf{D65} (2002),
  075024 [hep-ph/0110270]\relax
\relax
\bibitem{Buchmuller:2004rq}
W.~Buchm{\"u}ller, K.~Hamaguchi, M.~Ratz, T.~Yanagida, Phys. Lett.
  \textbf{B588} (2004), 90 [hep-ph/0402179]\relax
\relax
\bibitem{Brandenburg:2005he}
A.~Brandenburg, L.~Covi, K.~Hamaguchi, L.~Roszkowski, F.~D. Steffen, Phys.
  Lett. \textbf{B617} (2005), 99 [hep-ph/0501287]\relax
\relax
\end{mcbibliography}
\bibliographystyle{ArXivmcite}

\end{document}